\documentclass[a4paper,reprint,preprintnumbers,amsmath,amssymb,superscriptaddress]{revtex4-1}
\usepackage{cases}
\usepackage{txfonts}
\usepackage{color}
\usepackage{amssymb}
\usepackage{amsmath}
\usepackage{graphicx}
\usepackage{wrapfig}
\usepackage{dcolumn}
\usepackage{bm}
\usepackage{wasysym}
\usepackage{textcomp}
\usepackage{braket}
\usepackage{lineno}
\usepackage{soul,xcolor}
\usepackage{epstopdf}
\begin{document}
\title{Resonance fluorescence and laser spectroscopy of three-dimensionally confined excitons in monolayer WSe$_2$}
\author{S.~Kumar}%
	\affiliation{Institute of Photonics and Quantum Sciences, SUPA, Heriot-Watt University, Edinburgh EH14 4AS, UK}%
\author{M. Brotons-Gisbert}
	\affiliation{ICMUV, Instituto de Ciencia de Materiales, Universidad de Valencia, P.O. Box 22085, 46071 Valencia, Spain}
\author{R. Al-Khuzheyri}
	\affiliation{Institute of Photonics and Quantum Sciences, SUPA, Heriot-Watt University, Edinburgh EH14 4AS, UK}
\author{A.~Branny}
	\affiliation{Institute of Photonics and Quantum Sciences, SUPA, Heriot-Watt University, Edinburgh EH14 4AS, UK}
\author{G.~Ballesteros-Garcia}
	\affiliation{Institute of Photonics and Quantum Sciences, SUPA, Heriot-Watt University, Edinburgh EH14 4AS, UK}
\author{J. F. S\'{a}nchez-Royo}
	\affiliation{ICMUV, Instituto de Ciencia de Materiales, Universidad de Valencia, P.O. Box 22085, 46071 Valencia, Spain}
\author{B.~D.~Gerardot}%
	\email[e-mail: ]{B.D.Gerardot@hw.ac.uk}
	\affiliation{Institute of Photonics and Quantum Sciences, SUPA, Heriot-Watt University, Edinburgh EH14 4AS, UK}%
\date{\today}

\begin{abstract}
Resonant optical excitation of few-level quantum systems enables coherent quantum control, resonance fluorescence, and direct characterization of dephasing mechanisms. Experimental demonstrations have been achieved in a variety of atomic and solid-state systems. An alternative but intriguing quantum photonic platform is based on single layer transition metal chalcogenide semiconductors, which exhibit a direct band-gap with optically addressable exciton valley-pseudospins in a uniquely two-dimensional form. Here we perform resonance and near-resonance excitation of three-dimensionally confined excitons in monolayer WSe$_2$ to reveal near ideal single photon fluorescence with count rates up to 3 MHz and uncover a weakly-fluorescent exciton state $\sim$5\,meV blue-shifted from the ground-state exciton. We perform high-resolution photoluminescence excitation spectroscopy of the localized excitons, providing important information to unravel the precise nature of the quantum states. Successful demonstration of resonance fluorescence paves the way to probe the localized exciton coherence. Moreover, these results yield a route for investigations of the spin and valley coherence of confined excitons in two-dimensional semiconductors.
\end{abstract}

\maketitle

\section{Introduction}
Near-resonance optical excitation of quantized matter underpins the field of quantum photonics. It enables the initialization, coherent manipulation, and read-out of the quantum states~\cite{Press08,Ramsay08,Kim11,Poem11,Yale13,Xia15} and, via resonance fluorescence~\cite{Muller07,Wrigge08}, the generation of indistinguishable single photons~\cite{Lettow10,He13natnanotech,Sipahigil14,Proux15} - a crucial resource for future quantum technologies~\cite{Gao15}. In the solid-state, such quantum optical demonstrations have been made with quantum dots~\cite{Press08,Ramsay08,Kim11,Poem11,Muller07,Proux15}, single molecules~\cite{Wrigge08,Lettow10} and crystal defects~\cite{Yale13,Sipahigil14,Xia15}. For fundamental investigations, resonant or near-resonant optical excitation is invaluable  to probe the coherence and dephasing mechanisms in few-level quantum systems.

Rapid progress has recently been made in understanding the two-dimensional exciton (2D-$X$), spin, and valley-pseudospin properties in monolayer transition metal dichalcogenide semiconductors~\cite{Liu15,Xu14,Glazov15}. The first Brillouin zone of a monolayer transition metal dichalcogenide (TMD) such as MoS$_2$, WS$_2$, WSe$_2$, or MoSe$_2$ has a hexagonal shape that accommodates three pairs of degenerate but inequivalent edges, often denoted by $K$ and -$K$, which exhibit a direct band-gap with unique selection rules: for W-based TMDs left- (right-) handed circular polarized photons couple to interband transitions in the $K$ (-$K$) valley only ~\cite{Xu14,Glazov15,Jones13,Wang14}. Further, strong spin-orbit coupling links the spin and the valley-pseudospin, giving rise to spin-dependent optical selection rules. Also unique to these semiconductors with intrinsic two-dimensional confinement are very strong Coulomb interactions, large effective masses, and reduced dielectric screening which lead to large exciton binding energies ($\approx$0.5\,eV) and small Bohr radii ($<$1\,nm)~\cite{He14,Chernikov14}. 

Recently, localized excitons that exhibit substantially reduced linewidths compared to 2D-$X$ have been discovered in two-dimensional materials \cite{Srivastava15,He15,Koperski15,Chakraborty15,Tonndorf15,Kumar15,Tran16,Branny16}. However, besides their basic magneto-optical properties, these quantum emitters have yet to be explored in detail. Fundamental open questions revolve around the precise nature of the three-dimensional confinement and its effect on emitter properties, e.g. spin-orbit coupling and valley hybridization~\cite{Liu14}, which impact the potential for a coherent spin-valley qubit that can be coherently controlled with near-resonance optical excitation~\cite{Wu16}. Here we focus on single quantum emitters in WSe$_2$, in which a range of observed magneto-optical properties are broadly categorized as follows. (i) Emitters with a fine-structure splitting (FSS) of 0.6 to 0.8 meV caused by exchange interactions that exhibit a large (7 to 10) exciton g-factor ~\cite{Srivastava15,He15,Koperski15,Chakraborty15,Kumar15}. The FSS doublet typically exhibits equal intensity and orthogonal linear polarization but not exclusively~\cite{Kumar15}. (ii) Emitters with a smaller FSS doublet ($\approx$ 0.3 meV) with approximately parallel linear polarization and very small g-factor. (iii) Spectral lines without measurable FSS that do not exhibit any or extremely small Zeeman splitting even for $B_{\text{ext}}$\,=\,9\,T~\cite{He15}. These emitters are typically linearly polarized and the degree of linear polarization is unchanged with magnetic field. In this Letter we investigate emitters in categories (ii) and (iii).

\begin{figure} \centering
\includegraphics[width =83.668mm]{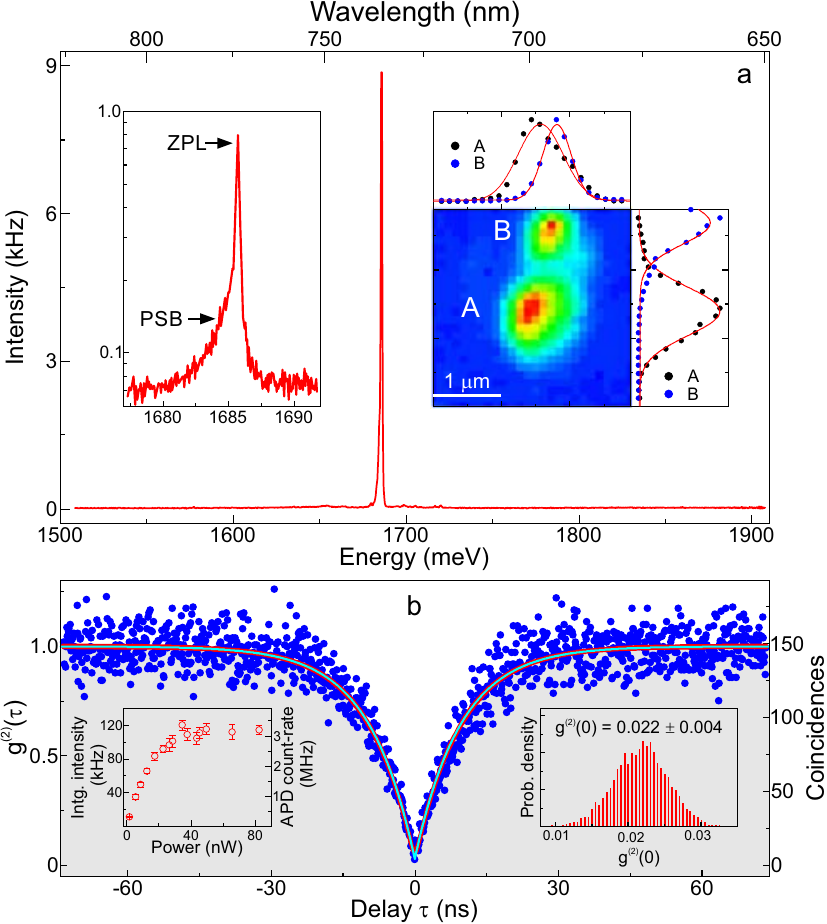}
\caption{\textbf{A highly isolated quantum emitter with high purity single photon emission.}~\textbf{a,}~A low-resolution emission spectrum from location A on the WSe$_2$ monolayer showing a single emission line from a single localized emitter. Inset left: A high-resolution spectrum showing the zero-phonon line (ZPL) and a low-energy phonon sideband (PSB). Inset right: Color-coded normalized peak intensities map of emitter A and a neighboring emitter B with strong spatial localization of both emitters.\textbf{b,}~Normalized second-order correlation function $g^{(2)}(\tau)$ of the emission line of emitter A exhibiting nearly perfect antibunching. The solid red line is a 95$\%$ confidence band for fitting of the measured data, yielding a deconvolved $g^{(2)}(0)$\,=\,0.022\,$\pm$\,0.004 and a lifetime of 9.55\,$\pm$\,0.11\,ns. The thin cyan line is the calculated deconvolved $g^{(2)}(\tau)$. Inset left: Power dependence of integrated intensity and photon counting rate for emitter A. Inset right: The probability density of $g^{(2)}(0)$ calculated using the probabilistic values of the fitted parameters. The most probable value of $g^{(2)}(0)$ and its standard deviation have been estimated from this plot. The size of time bin is 128\,ps. The measurements were performed using non-resonant CW excitation at $\lambda\,=\,532$\,nm with powers of 4\,nW for (a) and 2\,nW for (b).} \label{fig1}
\end{figure}

\begin{figure*} \centering
\includegraphics[width =168.21mm]{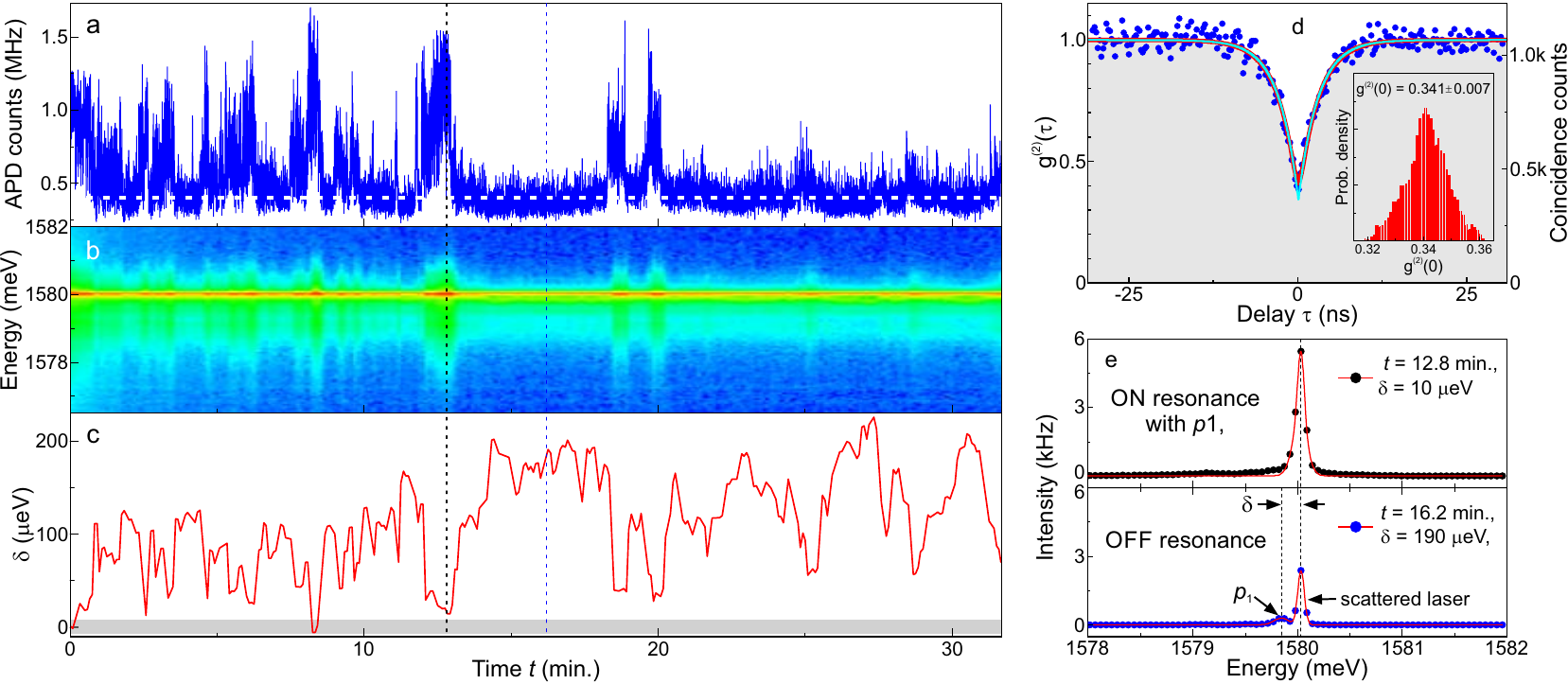}
\caption{\textbf{Resonance fluorescence from a single quantum emitter.}~Simultaneous time traces of the fluorescence from emitter B under resonant CW excitation at $\lambda\,=\,784.69460$\,nm with a power of 1\,\textmu W as recorded on~\textbf{a,}~an APD and~\textbf{b,}~a high-resolution spectrometer with 70\,ms and 5\,s integration, respectively. The background level of ~0.4\,MHz shown by a horizontal dashed line in (a) and the line at $\sim$1580.03\,meV in (b) are due to the scattered excitation laser.~\textbf{c,}~The time trace of emitter detuning $\delta\,=\,E_{\text{laser}}\,-\,E_{\text{p1}}$ of the dominant emission line $p_{\text{1}}$ of emitter B. The gray area is the fitting errors of $\delta$.~\textbf{d,}~ $g^{(2)}(\tau)$ for a time-interval when $\delta \approx 0$ showing that the total signal is antibunched. The solid thick line is a 95$\%$ confidence band for fitting of the measured data, yielding a deconvolved $g^{(2)}(0)$\,=\,0.341\,$\pm$\,0.007 and a decay-time of 2.87\,$\pm$\,0.05\,ns. The relatively high value of $g^{(2)}(0)$ is due to a signal-to-background ratio of $\sim$4.3~\textbf{e,}~The fluorescence spectra of emitter B at two different time instances marked by black (blue) dashed lines in (a-c) corresponding to time  $t$\,=\,12.8 (16.2)\,min. for $\delta$\,= 10 (190)\,\textmu eV. The black (blue) closed circles are measured data and solid lines are fits.} \label{fig2}
\end{figure*}

\begin{figure*} \centering
\includegraphics[width =168.98mm]{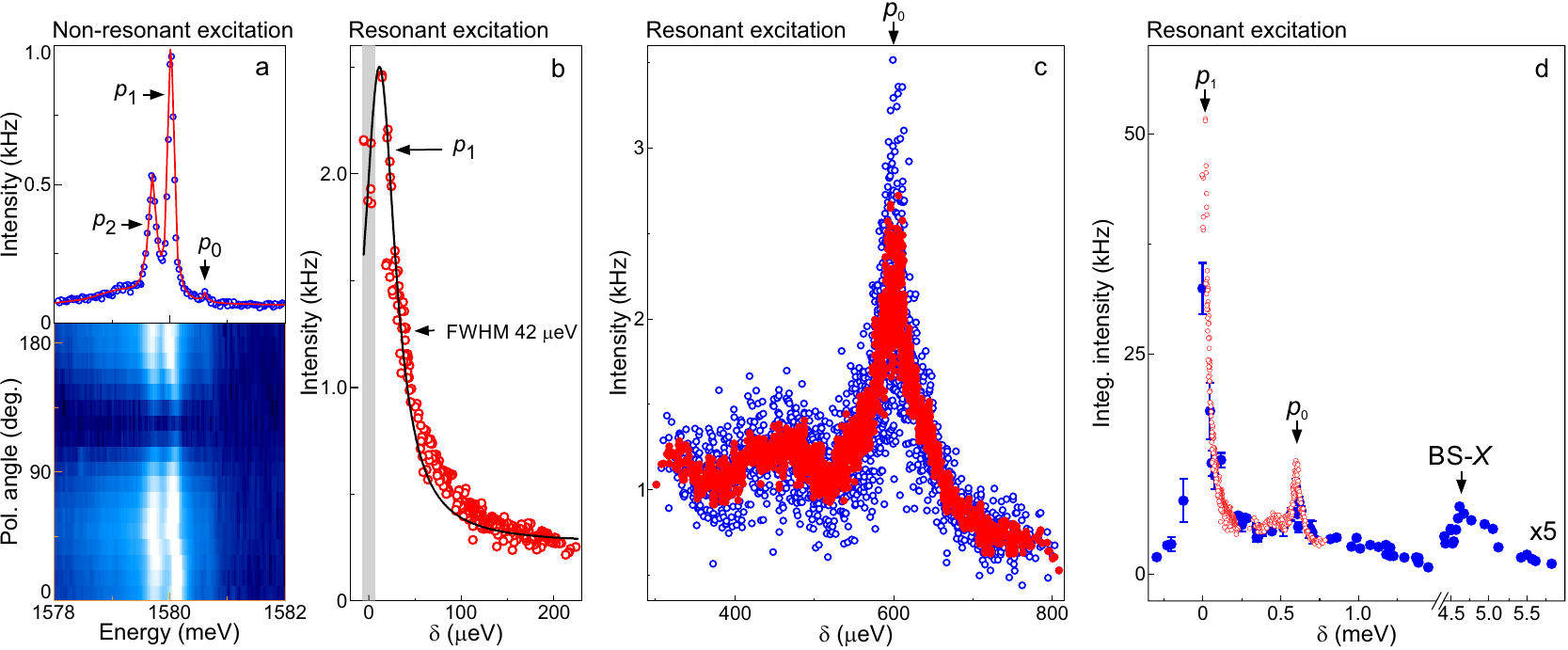}
\caption{\textbf{High-resolution PLE spectroscopy and observation of a weakly-fluorescent blue-shifted exciton (BS-$X$).}~\textbf{a,}~Polarization-resolved single fluorescence spectrum (top) and color-coded intensity map (bottom) of emitter B under non-resonant excitation showing three emission lines.~\textbf{b,}~A demonstration of high-resolution PLE spectroscopy. The fitted peak intensities (open circles) of the emission line $p_{\text{1}}$ in ~\ref{fig2}b as a function of $\delta$ identify $p_{\text{1}}$ beyond the resolution limit of the spectrometer and the spectral fluctuations. The gray area is the fitting errors of $\delta$.~\textbf{c,}~High-resolution PLE spectra showing the resonance for the line $p_{\text{0}}$. Open blue circles represent peak $p_{\text{1}}$ intensity as a function of $\delta$ and the closed red circles are the 5-point nearest-neighbour smoothed data. Two resonant excitation wavelengths (784.3800 and 784.4090\,nm) were used at a power of 10\,\textmu W and with a 50\,ms acquisition time.~\textbf{d,}~The PLE spectrum of emitter B shows the two bright-exciton peaks $p_{\text{0}}$ and $p_{\text{1}}$ and a BS-$X$ resonance. Closed blue circles are integrated intensities of the line $p_{\text{1}}$ obtained by scanning the laser wavelengths. The open red circles are PLE resonances for peaks $p_{\text{1}}$ and $p_{\text{0}}$. Peak $p_{\text{2}}$ is not visible in this experiment due to its lower emission energy.} \label{fig3}
\end{figure*}

\begin{figure} \centering
\includegraphics[width =89.796mm]{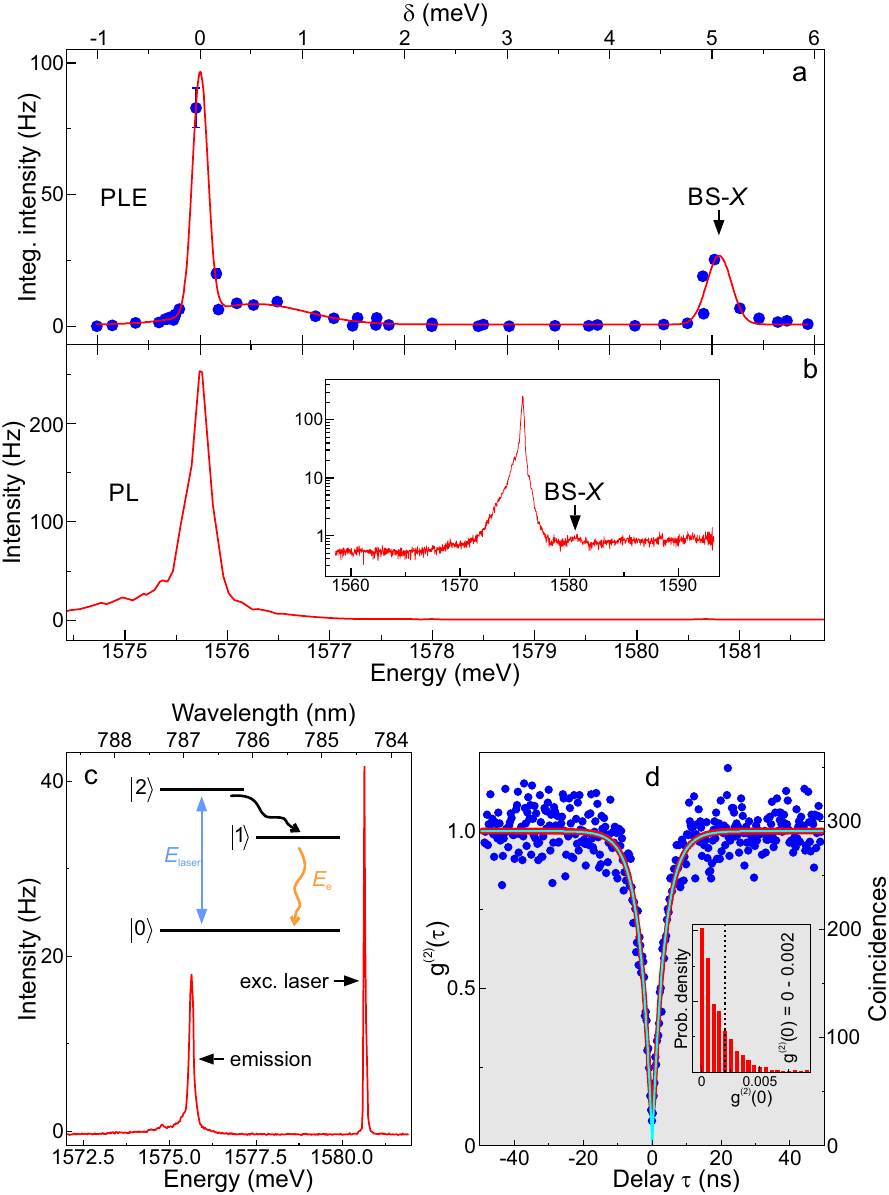}
\caption{\textbf{Emitter C: Observation of BS-$X$ in PLE spectroscopy and emission of high-purity single photons under resonant excitation of the BS-$X$.}~\textbf{a,}~PLE spectrum of emitter C identifying the ``BS-$X$" exciton at an energy $\sim$5\,meV higher than the ground-state exciton. The closed blue circles are data points while the solid red curve is guide to the eye composed of three Gaussian functions.~\textbf{b,}~The fluorescence spectrum of emitter C under non-resonant excitation at a power of 4\,\textmu W. The spectrum matches the energy range for which PLE is performed in (a). Insets: The full fluorescence spectrum of emitter C acquired on the high-resolution grating under non-resonant excitation. \textbf{c,}~ The fluorescence spectrum of emitter C under resonant excitation of the BS-$X$. The excitation power for (a) and (c) was 80\,nW.~Inset: Schematic of resonant excitation to the BS-$X$ and emission via the ground-state exciton.~\textbf{d,}~$g^{(2)}(\tau)$ under resonant CW excitation of the BS-$X$. High-purity single photon emission is observed with a deconvolved $g^{(2)}(0)$\,$<$\,0.002 and a decay-time of 3.50\,$\pm$\,0.05\,ns.}\label{fig4}
\end{figure}

\section{Results}


First, we show in Fig.~\ref{fig1} that monolayer WSe$_2$ is a suitable host for a pure single photon emitter. Under non-resonant excitation, a highly spectrally and spatially isolated emitter delivers single photon emission with a single photon purity $g^{\text{(2)}}(0)$\,$\approx$\,2$\%$ and a single photon count-rate $>$ 3\,MHz at saturation. A low-resolution microphotoluminescence (\textmu -PL) spectrum of emitter A,  described by emitter category (iii) above, is shown in Fig.~\ref{fig1}a. In contrast to all previous observations where sharp emission lines have been accompanied by extraneous emission from other localized emitters or 2D-$X$~\cite{Srivastava15,He15,Koperski15,Chakraborty15,Tonndorf15,Kumar15,Tran16,Branny16}, here we demonstrate an emission spectrum dominated by a single quantum emitter. The 2D-$X$ emission is highly suppressed as the optically excited electron-hole pairs are efficiently captured by a single confined exciton. The left inset of Fig.~\ref{fig1}a shows the high-resolution \textmu -PL spectrum, revealing the zero-phonon line (ZPL) and a low energy phonon sideband (PSB). The intensity ratio of ZPL:PSB is $\approx$\,60:40. Figure~\ref{fig1}b presents the second-order correlation function $g^\text{(2)}(\tau)$ under non-resonant CW excitation. Using Bayesian statistics (see Supplementary Sec.~III for details) to fit (solid lines) the measured data (closed circles), we obtain a deconvolved $g^{(2)}(0)$\,=\,0.022\,$\pm$\,0.004 (see also the right inset of Fig.~\ref{fig1}b for the probability density plot). This high single photon purity is essential for future quantum photonic applications.


We now present resonance fluorescence (RF) from a single quantum emitter in monolayer WSe$_2$. Emitter B, belonging to emitter category (ii) above, was chosen due to its favourable wavelength ($\lambda\,\approx$\,784.69\,nm) for our tunable laser diode and good spatial and spectral isolation (see right-inset of Fig.~\ref{fig1}a). Non-resonant \textmu -PL was first used to identify the ZPL wavelength and then the excitation laser was tuned into resonance. The background laser scattering was highly, but not completely, suppressed using orthogonal linear polarizers in the excitation and collection arms of the microscope (see Methods). We split the RF signal into two parts: 70$\%$  was measured by an APD (see Fig.~\ref{fig2}a) and 30$\%$ by a spectrometer (see Fig.~\ref{fig2}b). By fitting each spectrum in Fig.~\ref{fig2}b, the emitter peak energy detuning ($\delta\,=\,E_{\text{laser}}\,-\,E_{\text{p1}}$, where $E_{\text{p1}}$ is the peak emission energy), is determined (see Fig.~\ref{fig2}c). Two example spectra with fits are shown in Fig.~\ref{fig2}e. The single photon count-rate dynamics are directly correlated with $\delta$. We ascribe the slow spectral fluctuations to charge noise in the emitter environment, similar to that observed in semiconductor quantum dots~\cite{Houel12,Dekker91}. A maximum count-rate (mean background level) of $\sim$1.7 ($\sim$0.4)\,MHz is observed when the emitter is in-resonance (out-of-resonance) with the excitation laser. A second-order coherence measurement during a time-interval when the emitter was in resonance with the excitation laser yielded $g^{(2)}(0)$\,=\,0.341\,$\pm$\,0.007, conclusively demonstrating that the RF signal is indeed composed of quantum light (see Fig.~\ref{fig2}d). A signal-to-background of $\approx$ 4.3 is obtained by fitting the measured antibunching data, in agreement with the the maximum signal and background count-rates shown in Fig.~\ref{fig2}a.


The polarization properties of emitter B under non-resonant excitation are shown in Fig.~\ref{fig3}a. The brightest peak $p_{\text{1}}$ is accompanied by peak $p_{\text{2}}$ ($p_{\text{0}}$) on its low- (high-) energy side, energetically separated by $\sim$330 (600)\,\textmu eV. The polarization-resolved PL map (see Fig.~\ref{fig2}a: bottom) shows that peaks $p_{\text{1}}$ and $p_{\text{2}}$ are linearly polarized along almost the same direction and peak $p_{\text{0}}$ is polarized at slightly different angle. Notably, under resonant excitation conditions (see Fig.~\ref{fig2}b and e and Supplementary Figs.~S1 and S2), emission from $p_{\text{2}}$ is highly suppressed compared to non-resonant excitation. This result provides a hint that a specific valley-index can be optically addressed, encouraging further investigations.

We take advantage of the spectral fluctuations to perform high-resolution photoluminescence excitation (PLE) spectroscopy. This method allows us to measure several $\delta$ values with an accuracy of $\pm$5\,\textmu eV at a fixed excitation laser wavelength. Figure~\ref{fig3}b plots fitted peak intensities of peak $p_{\text{1}}$ versus $\delta$ (see open circles), both extracted from Fig.~\ref{fig2}b. It shows a resonance for peak $p_{\text{1}}$ with FWHM of 42\,\textmu eV. Similarly, PLE of peak $p_{\text{0}}$ is performed (see Fig.~\ref{fig3}c). The resonances of peaks $p_{\text{1}}$, $p_{\text{0}}$ and a high-energy phonon band are clearly resolved (see Fig.~\ref{fig3}d). More importantly, an additional resonance peak, blue-shifted by $\sim$4.75\,meV from $p_{\text{1}}$, is also observed. The PL spectrum of emitter B under non-resonant excitation shows negligible emission at this energy (see Supplementary Fig.~S1).\\
\indent To investigate if the blue-shifted exciton (BS-$X$) observed in PLE is an intrinsic property of quantum emitters in monolayer WSe$_2$, we probe a third emitter. Emitter C, from the same monolayer flake and belonging to emitter category (iii), exhibits a BS-$X$ detuned from the primary exciton by 5.07\,$\pm$0.01\,meV. We compare the PLE spectrum (Figs.~\ref{fig4}a) with a PL spectrum (Fig.~\ref{fig4}b, which has a high-resolution, logarithmic intensity scale spectrum shown in the inset). Compared to the ground-state exciton, the PL emission from BS-$X$ is suppressed by a factor of 1,250.


Finally, we establish high-purity single photon emission from the ground-state exciton under resonant-excitation of the BS-$X$ state. Figure~\ref{fig4}c shows a spectrum consisting of the low energy bright exciton emission and the scattered excitation laser due to imperfect polarization cancellation. This laser peak can be filtered with high fidelity, enabling clean $g^{(2)}(\tau)$ measurements as shown in Fig.~\ref{fig4}d. A deconvolved $g^{(2)}(0)$ value of $<$\,0.002 is achieved, demonstrating a single photon source with perfect purity. 

\section{Discussion}

The experiments have revealed that optical absorption into the BS-$X$ state quickly relaxes into the ground exciton state from which it can emit pure single photons. We currently do not understand the nature or origin of BS-$X$, but coupling to phonon modes can be excluded~\cite{Zhang15phonon}. One possibility is the BS-$X$ is a charged species that quickly relaxes into the ground state exciton, motivating future experiments with charge-tunable samples~\cite{Chakraborty15}. Another possibility is that the BS-$X$ state is a mostly optically inactive dark exciton. For W-based monolayer TMDs (WS$_2$, WSe$_2$), the electron spin in the lowest conduction band is antiparallel to carrier spin in the highest valence band, leading to optically inactive dark states which limits their quantum efficiency for light emission at low temperatures in comparison to Mo-based monolayer TMDs such as MoSe$_2$ ~\cite{Zhang15,Wang15}. Similar to valley hybridization, the optical activity of localized dark excitons could be linked to the symmetry of the confinement potential and underlying crystal lattice. Further investigations are required to understand the nature of the ground state excitons and BS-$X$. Tantalizingly, unlike with strict resonance fluorescence, the BS-$X$ offers future opportunities to investigate spin-valley coupling using excitation and fluorescence detection in both co-polarized or cross-polarized configurations.

In summary, we have demonstrated that monolayer WSe$_2$ is a benevolent host for a pure single photon emitter. These quantum emitters yield bright, stable, and highly-pure quantum light. The two-dimensional nature of the platform provides unique opportunities to engineer the light-matter interaction and integrate onto quantum photonic chips. We unambiguously achieve resonance fluorescence from the quantum emitters in spite of significant spectral fluctuations and background laser scattering. Strategies such as incorporating the single photon emitters into tunable electronic devices and surface passivation or encapsulation are likely to provide significant improvement. While the spectral fluctuations create challenges for quantum control and resonance fluorescence, we also demonstrate its utility for high-resolution PLE spectroscopy. PLE yields the direct observation of a three-dimensionally confined weakly-fluorescent exciton state that is energetically blue-shifted by $\sim$5\,meV. Resonant excitation of this BS-$X$ state provides an extremely robust and pure single photon source. The high-resolution characterization of the bright-exciton fine-structure and the experimental observation of the BS-$X$ are important results to better understand the specific nature of these localized excitons. The resonance fluorescence and laser spectroscopy techniques demonstrated here raise the prospect for indistinguishable single photon generation and investigations of the spin and valley coherence of strongly confined excitons in 2D-TMDs.

\section{Methods}
Using an all-dry viscoelastic stamping procedure~\cite{Castellanos-Gomez14}, we integrate a mechanically exfoliated WSe$_2$ flake onto a few layers of h-BN on top of a piezoelectric actuator so that in-plane dynamic strain could be induced in the flake by applying an out-of-plane electric field to the actuator. The actuator is made of a PMN-PT substrate. In the context of this letter, all experiments have been performed at zero external electric field to the actuator, and therefore, both top and bottom Ti/Au (5/100\,nm) electrodes of the actuator have been grounded. All measurements have been performed on a single monolayer, which has been identified using optical micrographs and spatial maps of \textmu -PL.

A confocal microscope with an objective lens with NA of 0.82, yielding a diffraction limited focus of $\sim$ 460 nm at $\lambda$ = 750 nm, was used for resonant laser spectroscopy. A CW tunable laser diode, covering a wavelengths range of 765 - 805\,nm, was used for resonant excitation. $\lambda$\,=\,532\,nm was used for non-resonant CW excitation. The fluorescence signal was separated from the excitation laser via orthogonally oriented linear polarizers in the excitation and collection arms of the microscope. This yields a 10$^7$ suppression of laser counts on smooth substrates, but the rough gold surface used here yields 10$^5$ suppression at best. The sample was placed on automated nanopositioners at T = 4\,K in a closed-cycle cryostat.  All spectra were acquired with a 0.5 m focal length spectrometer and nitrogen-cooled charge-coupled device with a measured spectral resolution of $\sim$75 \textmu eV at $\lambda$\,=\,784\,nm for an 1800 l/mm grating. A separate confocal microscope is used to perform the polarization-resolved \textmu -PL measurements. A fiber based Hanbury-Brown and Twiss interferometer was used for second-order correlation measurements and photon counting was performed using Si avalanche photodiodes.

\begin{acknowledgments}
We thank B. Urbaszek for fruitful discussion, A. Rastelli for data analysis software, and A.C. Dada for assistance with the experimental setup. This work was supported by a Royal Society University Research Fellowship, the EPSRC (grant numbers EP/I023186/1, EP/K015338/1, and EP/L015110/1) and an ERC Starting Grant (number 307392), and ESP Grants. Support from the Spanish Government (Grant No. TEC2014-53727-C2-1-R), Comunidad Valenciana Government (Grant No. PROMETEOII/2014/059), and University of Valencia (UV-INV-PREDOC13-110538) is acknowledged.\\
\end{acknowledgments}






\end{document}